# Tunable optical multistability induced by a single cavity mode in cavity quantum electrodynamics system

**Liyong Wang[1,*] , Yinxue Zhao[2] and Jiajia Du[3,*]**

[1]Department of Applied Physics, Wuhan University of Science and Technology, Wuhan 430081, China

[2]Wuhan Social Work Polytechnic, Wuhan 430079, China

[3]Chongqing University of Post and Telecommunications, Chongqing 400065, China

E-mail: **wangliyong@wust.edu.cn; dujj@cqupt.edu.cn**

**Abstract.** A tunable optical multistability scheme based on a single cavity mode coupled with two separate atomic transitions in an atom-cavity system is proposed and demonstrated. Under the collective strong coupling condition, multiple polariton eigenstates of the atom-cavity system are produced. The threshold and optical multistability curve can be tuned freely by system parameters in a broadband range. Moreover, a certain bistability region of the system is split to two bistability regions due to destructive quantum interference induced by an extra weak control field. Compared to traditional optical multistabilities created by two or more light fields, the proposed optical multistability scheme has compactness and is easy to be miniaturized. The proposed scheme is useful for manufacturing integrated application of multi-state all-optical logic devices and constructing basic elements of all-optical communication networks.

## 1. Introduction

Over the past decades, optical bistability and optical multistability which mainly originate from the nonlinearities of optical media induced by light fields in resonantors, have been widely investigated theoretically and experimentally due to their potential applications in all-optical switching [1-4], all-optical coding elements [5, 6], optical memories [7-10], quantum information processing [11-14]. Optical multistabilities usually exist in two forms: one is that a certain input light intensity value leads to three or more output light intensity values, and the other is multiple bistability regions existing simultaneously in an input-output curve of a physical system. With exponential increase of transmission data in the future, it is an essential requirement that all-optical logical devices should work in parallel with multiple channels to transport and transfer data in all-optical computation and quantum teleportation networks [15, 16]. Optical multistability can be used to realize multistate quantum elements like all-optical switching [17], all-optical coding and routing [18], all-optical logic gates [19, 20], all-optical transistors [5, 21], etc. Compared to optical bistability, optical multistability is more difficult to obtain since it usually requires high-order nonlinearity or multiple nonlinearity regions in the absorptive or dispersive media. Traditional optical multistability can be realized with $\Lambda$-type [22] or V-type [23] three-level atom-cavity systems, and it requires large input field detuning to create high-order nonlinearity [24]. Optical tristabilities based on degenerate Zeeman sublevels created by extra magnetic fields are also reported, and the performances of tristabilities can be tuned by extra magnetic fields [25-27]. Besides, optical multistabilities are proposed to be obtained with more than two light fields interacting with multi-energy-level atoms of diamond type configuration [28], N-type configuration [29], Y-type configuration [30], etc. These schemes enhance the tunability of optical multistability due to the quantum coherence and interference in multi-energy-level atomic systems, but the extra light fields inevitably increase the complexity of systems and make them difficult to miniaturize. Optical multistabilities based on spontaneously generated coherence (SGC) effect are also reported [31, 32]. However, the requirements of nonorthogonal dipole matrix elements and near-degenerate energy-levels are usually difficult to be realized in practical physical systems [29, 33]. In the year of 2015, a single cavity mode coupled with two separate atomic transitions in an atom-cavity system was realized experimentally. There are three tunable polariton rersonant peaks in the output spectrum, which is very different from the traditional normal mode spillting in the two-level atom-cavity system [34]. In 2019, a single cavity mode coupled with three separate atomic transitions and its output spectrum are also investigated experimentally [35], which further renders subtle quantum control and manipulation behaviours of light-matter interactions.

In this paper, we propose a new kind of optical multistability scheme based on a single optical cavity mode coupling with atoms in a cavity quantum electrodynamics (CQED) system. Two separate atomic transitions are excited simultaneously by a single cavity mode under the collective strong coupling condition. Optical multistability is produced by a single input optical signal field. The nonlinear input-output curve and thresholds (the injected optical field intensity which corresponds to the switching point of the output optical field intensity) can be tuned flexibly by system parameters. The proposed optical multistability scheme has relaxed configuration and may be designed as a multi-state optical passive device, thus it will be useful for applications like constructing multi-state quantum logic devices [36], multi-state all-optical switching [37-39], etc.

## 2. Theoretical Model

Fig. 1(a) shows the schematic setup. An unidirectional optical ring cavity consists of four cavity mirrors $M_i$ (i = 1-4). The reflectivities of mirrors $M_3$ and $M_4$ are 1. R and T denote the reflection and transmission coefficients of mirrors $M_1$ and $M_2$. R+T = 1. An input signal field $E_{\text{in}}^{s}$ is injected into the cavity from the left side of cavity mirror $M_1$. The single cavity mode circulates in the ring cavity and interacts with the atom ensemble inside the optical cavity [38, 39], then the output field $E_T^s$ comes out from the right side of cavity mirror $M_2$. A free-space control field $E_c$ interacts with the atom ensemble in the optical cavity. The CQED system is excited in nonlinear regime when the optical pumping effect induced by the signal field can not be neglected, and the population of atoms in excited states are nonzero [40, 41].

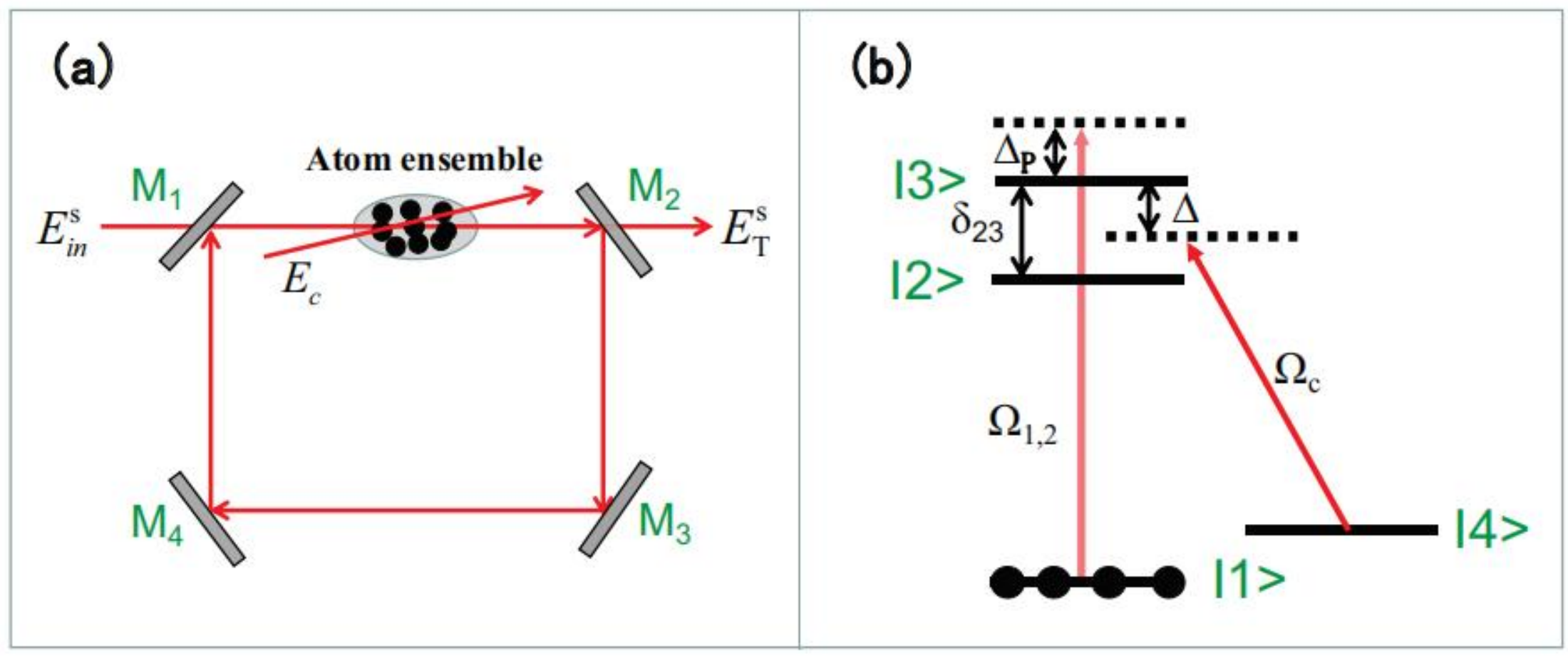

Figure 1. Schematic diagram of optical multistability created by a single optical cavity mode in CQED system. (a) An unidirectional optical ring cavity containing an atomic ensemble. An signal field is coupled into the cavity from cavity mirror $M_1$ and triggers optical cavity mode. (b) Atomic energy-level configuration for manipulation of a single cavity mode coupling with two separate atomic transitions.

As shown in Fig. 1(b), a single cavity mode couples with three energy-levels of atoms under the collective strong coupling condition [40, 42, 43]. The input signal field $E^s_{\text{in}}$ is detuned from atomic transition $|1\rangle \rightarrow |3\rangle$ by $\Delta_p = \omega_p - \omega_{31}$. The cavity mode detuning is defined as $\Delta_c = \omega_{cav} - \omega_{31}$. Two separate atomic transitions $|1\rangle \rightarrow |2\rangle$ and $|1\rangle \rightarrow |3\rangle$ with Rabi frequency $\Omega_1$ and $\Omega_2$ are excited simultaneously by the single cavity mode. $\delta_{23}$ is the frequency separation of two atomic excited states $|2\rangle$ and $|3\rangle$. A weak control field $E_c$ with Rabi frequency $\Omega_c$ couples the atomic transition $|4\rangle \rightarrow |3\rangle$ and the control field detuning is defined as $\Delta = \omega_c - \omega_{34}$. The Rabi frequencies are defined as $\Omega_1 = \mu_{12} E^s_{in} / (2\hbar)$, $\Omega_2 = \mu_{13} E^s_{in} / (2\hbar)$ and $\Omega_c = \mu_{43} E^s_{in} / (2\hbar)$. $\sigma^{(i)}_{mn}$ (m, n=1-4) is the atomic operator for the *i*th atom. The interaction Hamiltonian of the CQED system in this case is given as:

$$H = -\hbar\left(\Omega_c \hat{\sigma}_{34} + \Omega_1 \hat{\sigma}_{21} + \Omega_2 \hat{\sigma}_{31}\right) + H.C. \quad (1)$$

Under the rotating wave approximation framework, the system equations of motion for atoms are solved by $\frac{d\hat{\rho}}{dt} = \frac{1}{i\hbar}\left[\hat{H}, \hat{\rho}\right] + \hat{L}\hat{\rho}$ [44, 45]. L denotes the quantum superoperator of dissipation. The system equations of motion for atoms are obtained as:

$$\dot{\sigma}_{11} = \gamma_{21}\sigma_{22} + \gamma_{31}\sigma_{33} + \gamma_{41}\sigma_{44} + i\Omega_1^*\sigma_{21} - i\Omega_1\sigma_{12} + i\Omega_2^*\sigma_{31} - i\Omega_2\sigma_{13}, \quad (2)$$

$$\dot{\sigma}_{22} = -\gamma_2\sigma_{22} + i\Omega_1\sigma_{12} - i\Omega_1^* a\sigma_{21}, \quad (3)$$

$$\dot{\sigma}_{33} = -\gamma_3\sigma_{33} + ig_2 a\sigma_{13} - ig_2 a^+\sigma_{31} + i\Omega_c\sigma_{43} - i\Omega_c^*\sigma_{34}, \quad (4)$$

$$\dot{\sigma}_{44} = \gamma_{24}\sigma_{22} + \gamma_{34}\sigma_{33} - \gamma_{41}\sigma_{44} + i\Omega_c^*\sigma_{34} - i\Omega_c\sigma_{43}, \quad (5)$$

$$\dot{\sigma}_{23} = -[(\gamma_2 + \gamma_3)/2 - i\delta_{23}]\sigma_{23} + i\Omega_1\sigma_{13} - i\Omega_2^*\sigma_{21} - i\Omega_c^*\sigma_{24}, \quad (6)$$

$$\dot{\sigma}_{42} = -[\gamma_2/2 + i(\Delta - \Delta_p)]\sigma_{42} - i\Omega_1^*\sigma_{31} + i\Omega_c^*\sigma_{32}, \quad (7)$$

$$\dot{\sigma}_{43} = -(\gamma_3/2 + i\Delta)\sigma_{43} + i\Omega_c^*(\sigma_{33} - \sigma_{44}) - i\Omega_2^*\sigma_{41}, \quad (8)$$

$$\dot{\sigma}_{12} = -[\gamma_2/2 + i(\Delta_p + \delta_{23})]\sigma_{12} + i\Omega_1^*(\sigma_{22} - \sigma_{11}) + i\Omega_2^*\sigma_{32}, \quad (9)$$

$$\dot{\sigma}_{13} = -(\gamma_3/2 + i\Delta_p)\sigma_{13} + i\Omega_2^*(\sigma_{33} - \sigma_{11}) + i\Omega_1^*\sigma_{23} + i\Omega_c^*\sigma_{14}, \quad (10)$$

$$\dot{\sigma}_{14} = -[\gamma_{41}/2 + i(\Delta - \Delta_p)]\sigma_{14} + i\Omega_1^*\sigma_{24} + i\Omega_2^*\sigma_{34} - i\Omega_c\sigma_{13}, \quad (11)$$

where $\gamma_i$ denotes spontaneous decay rate of atom from upper state $|i\rangle$ to ground state $|1\rangle$. $\gamma_{41}$ denotes spontaneous decay rate between atomic states $|4\rangle$ and $|1\rangle$. Consider a closed atom system, thus $\sigma_{11}+\sigma_{22}+\sigma_{33}+\sigma_{44}=1$. Under the slowly varying envelope approximation, the dynamic equation of signal field governed by Maxwell's equation is:

$$\frac{\partial E^s}{\partial t} + c\frac{\partial E^s}{\partial z} = i\frac{\omega_p}{2\varepsilon_0}P\left(\omega_p\right) \quad (12)$$

where $P\left(\omega_p\right)=N(\mu_{12}\sigma_{12}+\mu_{13}\sigma_{13})$ is the induced polarization of three-level atomic excitation by a single cavity mode. The dipole matrix elements for atomic transitions $|1\rangle \to |2\rangle$ and $|1\rangle \to |3\rangle$ are assumed to be identical, i.e., $\mu_{12}=\mu_{13}=\mu$. N is the effective number density of atoms in the cavity. $\varepsilon_0$ is permittivity of free space and *c* is the light speed in vacuum. In steady state, the left sides of Eqs. (2-11) are zero. Then the boundary conditions of input field $E_{in}^s$ and output field $E_T^s$ for the CQED system are [12]:

$$E^s(0) = \sqrt{T}E_{in}^s + RE^s(\mathrm{L}) \quad (13)$$

$$E_T^s = \sqrt{T}E^s(\mathrm{L}) \quad (14)$$

L is the length of atomic ensemble. Solving Eqs. (12-14) under the mean-field limit [22, 23], the input-output relationship of the CQED system is given as:

$$y = 2x - i(\mathrm{C}_1\sigma_{12}+\mathrm{C}_2\sigma_{13}) \quad (15)$$

where $C_1 = LN\omega_p\mu_{12}^2/2\hbar c\varepsilon_0 \mathrm{T}$ and $C_2 = LN\omega_p\mu_{13}^2/2\hbar c\varepsilon_0 \mathrm{T}$ are the cooperativity parameters.

$C_1$=$C_2$=C since the dipole matrix elements $\mu_{12}$ and $\mu_{13}$ are assumed to be identical. $x = \mu E_T^s / (2\hbar\sqrt{T})$ and $y = \mu E_{in}^s / (2\hbar\sqrt{T})$. The input field intensity $I_{in} = |y|^2$ and the output field intensity $I_T = |x|^2$ can be obtained from Eq. (15).

## 3. Results

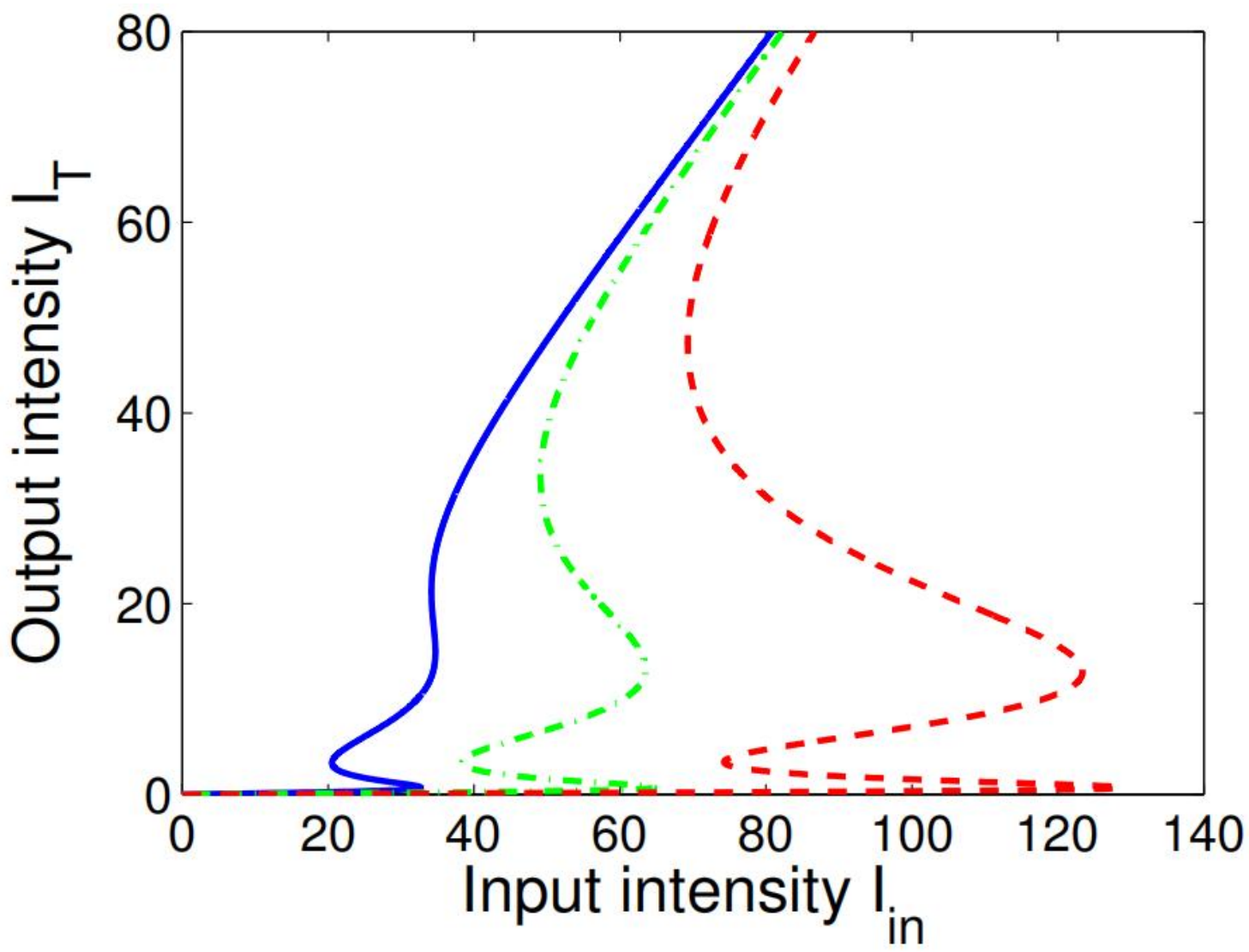

Figure 2. The input-output curve of the CQED system for single cavity mode coupling with three-level atoms without the control field ( Ωc=0). The blue full, green dash-dotted and red dashed lines correspond to C=90, 180 and 380, respectively. Other parameters are $\delta_{23}$=$12\Gamma$ , $\Delta_p$=0.

Different from the traditional normal mode splitting in two-level atom-cavity system, two separate atomic transitions $|1\rangle \rightarrow |2\rangle$ and $|1\rangle \rightarrow |3\rangle$ are excited simultaneously due to the collective strong coupling of atoms with a single cavity mode. Fig. 2 shows the nonlinear input-output relation of the CQED system when a single cavity mode excites three-level atoms. Optical tribistability is produced when the control light is absent (Ωc=0). As the cooperative parameter *C* increases, both the upper and lower thresholds of the optical tribistability increase. The lower threshold increases fastly than the upper threshold, thus the tribistability area is enlarged greatly. The tribistability curve also can be influenced by the energy-level separation $\delta_{23}$ of two upper atomic states, the input field detuning $\Delta_p$, etc. Fig. 3(a) shows tribistability curve as the energy-level separation $\delta_{23}$ of two upper atomic states decreases. The lower thresholds increases while the upper threshold decreases. As a result,

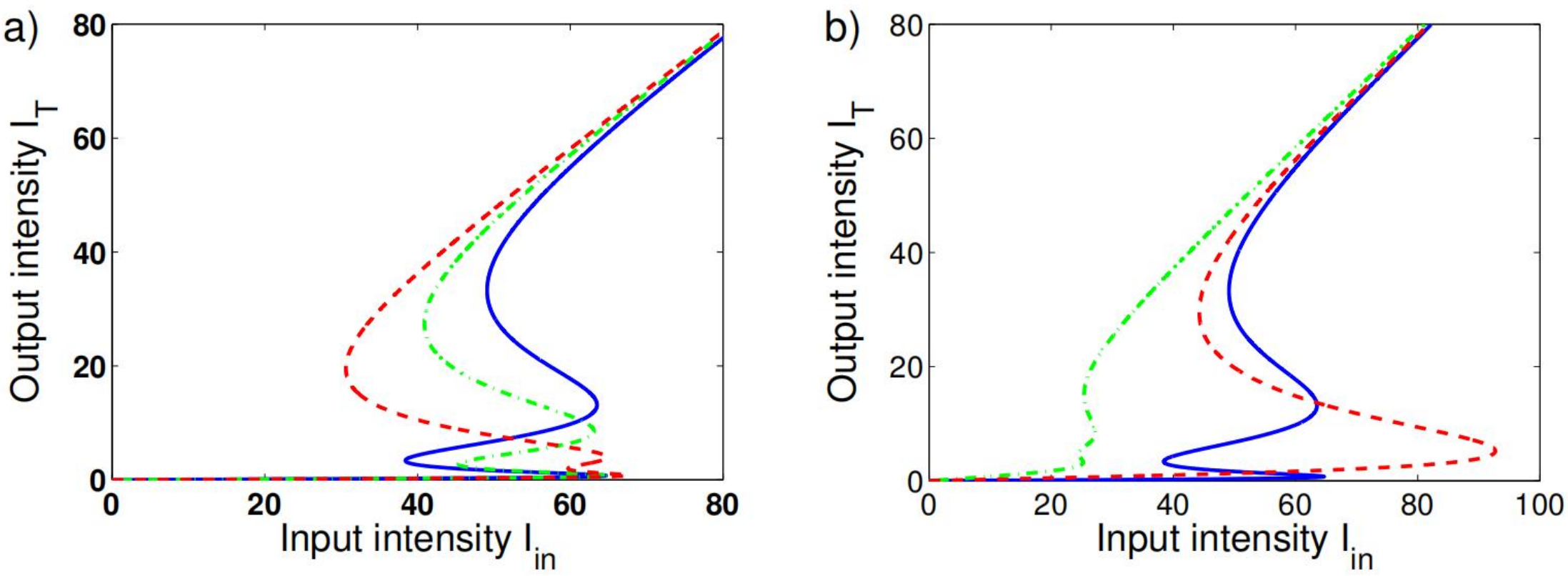

Figure 3. The input-output curve of a single cavity mode coupling with three-level atoms in the CQED system without the control field ($\Omega$c=0). (a) The output light intensity $I_T$ versus the input light intensity $I_{in}$. $\Delta_p$ =0. The blue full, green dash-dotted and red dashed lines correspond to $\delta_{23}$=12$\Gamma$, 8$\Gamma$, 4$\Gamma$, respectively. (b) The output light intensity $I_T$ versus the input light intensity $I_{in}$. $\delta_{23}$=12$\Gamma$. The blue full, green dashed dotted and red dashed lines correspond to $\Delta_p$ =0, -3$\Gamma$, 3$\Gamma$, respectively.

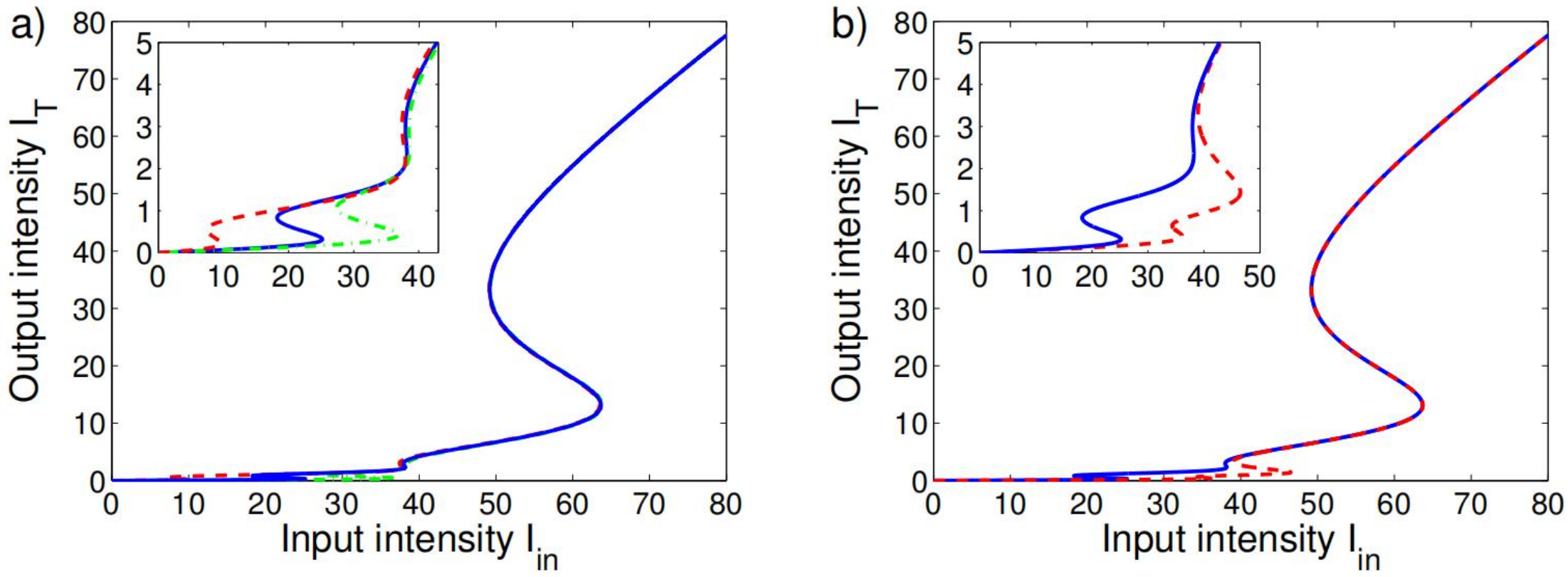

Figure 4. The input-output curve of a single cavity mode coupling with two separate atomic transitions in the CQED system with a free-space control field. $\delta_{23}$=12$\Gamma$, $\Delta$=0. (a) The variation of multi-bistable curve with the input field detuning $\Delta_p$. The red dashed, blue full and green dash-dotted lines correspond to $\Delta_p$=0.01$\Gamma$, 0.03$\Gamma$ and 0.05$\Gamma$, respectively. $\Omega_c$=0.1$\Gamma$. (b) The variation of multi-bistable curve with the control field intensity $\Omega_c$. $\Delta_p$=0.03$\Gamma$. The red dashed and blue full dotted lines correspond to $\Omega_c$=0.06$\Gamma$, 0.1$\Gamma$, respectively.

the lower bistability area decreases and the upper bistability area increases. Fig. 3(b) shows the tribistability curve with varying input field detuning $\Delta_p$. The tribistability areas enlarges as the input field frequency is tuned to the middle of two upper atomic states. That is because the excitation of two atomic transitions by the single cavity mode is enhanced. The thresholds of the tribistability also increase as the absorption of atomic medium to cavity field is strengthened. However, the nonlinear input-output curve is transferred from tribistable to bistability as the $\Delta_p$ is tuned away from two upper excited state, as is shown by the red dashed line in Fig. 3(b). In this case, the transition $|1\rangle \rightarrow |2\rangle$ can be neglected as the input field detuning $\Delta_p$ is large enough, and the CQED system acts as a two-level atom-cavity system.

For the multiple atomic energy-level excitation by a single cavity mode in the CQED system, destructive quantum interference can be created by an extra weak control field injected from the free space of the optical cavity [44]. The control field $E_c$ only couples the atomic transition $|4\rangle \rightarrow |3\rangle$. An electromagnetic-induced-transparency-like dip occurs at certain polariton state when the double resonance condition $\Delta = \Delta_p$ is satisfied. Fig. 4 shows input-output relation of four-level atom excited with a single cavity mode and a control field (see Fig. 1(b)). In Fig. 4(a), the control field frequency is tuned to be resonant to the atomic transition $|4\rangle \rightarrow |3\rangle$. Due to the destructive quantum interference induced by the control field, the original lower bistability region in Fig. 3 is split to two new bistability regions, and there are three bistability regions in the input-output curve of the CQED system. Thus a broadband multi-throw all-optical switching can be designed which is the fundamental device in all-optical communication [47, 48], quantum information processing [11, 12], etc. In Fig. 4(a), as the input field detuning $\Delta_p$ increases, the threshold and area of the lower bistability region increase. However, the upper bistability curve keeps unchanged. That is to say, the destructive quantum interference changes the nonlinear properties of the system near the polariton state when the double resonance condition is satisfied, while the nonlinearity of the other frequency ranges of the system are not affected, which is also demonstrated in Ref. [44]. In Fig. 4(b), the threshold of the lower bistability region decreases as the control field intensity $\Omega$ increases, which attributes to the decrease of absorption in atomic medium to the signal field. Based on this, a broadband three-state all-optical switching can be designed and manipulated in a wide operating range.

## 4. Conclusion

To summarize, a broadband and tunable optical multistability scheme based on a single cavity mode in CQED system is demonstrated. Tristability and multi-bistability are showed in the input-output curves, and their thresholds can be tuned flexibly by varying the system parameters. Due to the destructive quantum interference induced by an extra weak control field, the original bistability region can be split to two new bistability regions, which may be designed as a broadband multi-throw all-optical switching. The proposed scheme can be realized experimentally in many physical systems with moderate system parameter requirement [47, 48]. For a practical example of two separate atomic transitions excited by a single cavity mode, the signal field couples atomic ground state $\left|{}^2S_{1/2},\ F=2\right\rangle$ and two upper excited states $\left|{}^2P_{3/2},\ F=1\right\rangle$ and $\left|{}^2P_{3/2},\ F=2\right\rangle$ of $^{85}$Rb $D_2$ line; A weak control field couples atomic ground state $\left|{}^2S_{1/2},\ F=3\right\rangle$ and excited state $\left|{}^2P_{3/2},\ F=2\right\rangle$ of $^{85}$Rb $D_2$ line. For three separate atomic transitions excited by a single cavity mode, the signal field couples atomic ground state $\left|{}^2S_{1/2},\ F=2\right\rangle$ and three upper excited states $\left|{}^2P_{3/2},\ F=1\right\rangle$, $\left|{}^2P_{3/2},\ F=2\right\rangle$ and $\left|{}^2P_{3/2},\ F=3\right\rangle$ of $^{85}$Rb $D_2$ line. Compared to the traditional optical multistability phenomena which are usually produced by two or more light fields (or magnetic fields) in optical resonantors, the proposed optical multistability scheme based on a single cavity mode coupling with multi-level atoms in a collective strong coupling CQED system is easy to be miniaturized and cascaded due to the compactness [51, 52], which is the key requirement for integrated manufacturing applications. The proposed optical multistability scheme may also be desiged as a multistate passive optical device which is useful for applications like multistate all-optical switching [53, 54, 55], all-optical memory [54, 55], all-optical communications [47, 48], all-optical quantum logic elements [58], etc.

## Acknowledgments

This work was supported by the National Natural Science Foundation of China (Grant No. 12304292) and the Postdoctoral Applied Innovation Program of Shandong (Grant No. 62350070311227).

## Conflict of interest

The authors declare that there are no conflicts of interest related to this article.

## Data availability statement

The data that support the findings of this study are available upon reasonable request from the authors.